\documentstyle{article}
\begin{document}

\title{Some Key Issues Confronting Inflationary Cosmology}

\author{Robert H. Brandenberger \cite{conf} 
\\Physics Department, Brown University, Providence, RI. 02912, USA.
\\rhb@het.brown.edu}

\maketitle

\begin{abstract}
Does inflation provide a compelling explanation for why the universe is so large, so flat, and so old, and a predictive theory of density perturbations? In this brief contribution (based on the role of the author as moderator of the discussion session on inflation), a list of some of the key issues confronting inflationary cosmology will be given, with the hope of focusing the debate on inflation and drawing more attention to some of the potential problems of the inflationary theory.
\end{abstract}

\vskip1.cm
\noindent BROWN-HET-1056

\noindent August 1996
\vskip1.cm

Over the past fifteen years, the inflationary scenario \cite{original,reviews} has evolved into the most widely accepted theory of the very early Universe. Many physicists and astronomers in fact go as far as to consider inflation to be part of standard cosmology. As discussed by Alan Guth in these proceedings, it is asserted that inflation solves many of the problems of standard Big Bang cosmology, e.g. the horizon and flatness problems. Inflation may also lead to a simple mechanism of structure formation \cite{structure}. The seeds for the density fluctuations responsible for explaining CMB anisotropies and generating inhomogeneities on scales which at the present time are cosmological, originate as quantum fluctuations which are produced during the epoch of inflation on scales smaller than the Hubble radius and which are stretched to cosmological super-Hubble radius distances by the exponential expansion of the Universe. These fluctuations can be described by a Gaussian random field, which makes the analy

sis of structure formation tractable.

However, inflation is based on extrapolating known physical theories to energy scales much larger than can be probed experimentally in a laboratory. Therefore, it is important to carefully scrutinize the foundations of inflationary cosmology, and to isolate clear quantitative predictions which will allow the theory to be falsified or verified. Without falsifiability, the inflationary scenario cannot be considered as a complete scientific theory, much less part of the standard model.

I wish to put forward a partial list of {\it key issues} confronting inflation, in the hope of generating discussion and pointing to topics which need to be investigated in greater depth.

1. {\it Does inflation live up to its promise of solving the horizon and flatness problems?} This issue was debated at length by Bill Unruh and Alan Guth, and I refer to their contributions. A followup question is: {\it Are there other viable solutions of the homogeneity and flatness problems?} It is often stated that inflation is the only known solution to these problems. However, this is not true. An oscillating universe model can naturally explain the homogeneity problem, as discussed recently in Ref. \cite{durrer}. There may be other solutions, and this issue deserves more attention. 

2. {\it Does it make sense to speak about inflation driven by a temporary cosmological constant when we do not know how to solve the cosmological constant problem? In particular, does it make sense to use fundamental scalar fields to generate inflation since the zero of the scalar field potential energy is arbitrary?} An easy answer to this question is to say that we can analyze the gravitational dynamics of the solar system quite successfully if we set the cosmological constant $\Lambda$ equal to zero by hand, and that we should in the same way be able to set $\Lambda$ to zero when studying the gravitational dynamics of the early universe. However, there is a major difference as soon as we discuss inflation. Unlike in solar solar system dynamics which is {\bf not} driven by $\Lambda$, inflation is driven by a temporary $\Lambda$, i.e. the part of the theory which is not understood. Hence it is illegitimate to remove $\Lambda$ by hand. My view on this issue is that a convincing realization of inflation is st

ill lacking, and that such a realization should probably be based on a new fundamental principle (as discussed e.g. in Ref. \cite{nonsingular}) rather than on special features of an ad-hoc scalar field potential.

3. {\it Do we have a good model for inflation, well-motivated by some microphysical theory for which there is independent confirmation? Why are the fluctuations in this model small enough to be compatible with CMB anisotropies and with structure formation considerations? Is any fine-tuning required?} The second half of this question is the famous ``fluctuation problem" which has occupied cosmologists working on inflation since 1982 \cite{structure}. Many of the models discussed by particle physicists in recent years are natural \cite{natural} in the technical sense that small numbers remain small when quantum corrections are taken into account at each order in perturbation theory. However, there is still the need to have ad-hoc small numbers in the basic Lagrangian. Recently, there have been many attempts to construct models of inflation in which the small numbers required for inflation are coupled to small numbers which must be contained in any particle physics model in order to explain the hierarchy proble

m, e.g. the small ratio of Yukawa couplings required to explain the electron and the proton masses. Most successful seem supersymmetric models (see e.g. Ref. \cite{lyth} and references contained therein).

4. {\it Is inflation falsifiable?} If inflation is to become a scientific ``theory" rather than simply a ``scenario", the answer to this question must be ``yes". For a long time it was believed that $\Omega = 1$ is a prediction of inflation. However, counterexamples of inflationary models which predict $\Omega \neq 1$ have been known from the early days of inflation (see e.g. Ref. \cite{ellis}) and recently many models of inflation predicting $\Omega \neq 1$ have been constructed (see e.g. Ref. \cite{open}). As emphasized by Andy Albrecht in his contribution, there is a sub-issue which can be raised: {\it Is the idea that fluctuations from inflation are responsible for the observed structures in the universe falsifiable?} There has recently been a lot of progress towards answering this sub-question in the affirmative \cite{andy}. If adiabatic perturbations from inflation are responsible for the observed structure in the Universe, then the power spectrum of CMB anisotropies is expected to show characteristic 

acoustic oscillations. If these prove to be absent, this would provide strong evidence against inflation as the seeds of structure formation. Note that it is quite possible that inflation took place in the very early universe, but that the resulting perturbations were too small to be important today, and that another mechanism - such as topological defects \cite{defects} - is responsible for generating inhomogeneities. However, we are still left with the problem of coming up with an observational criterion with which the idea that inflation took place in the very early universe can be falsified.

5. {\it Is there evidence in favor of inflation from observations?} At first glance, this question may sound heretical. What about the beautiful agreement between the amplitude of CMB anisotropies on COBE scales and the power spectrum of density perturbations? This is in fact good evidence for a nearly scale-invariant spectrum of primordial density perturbations, precisely the spectrum predicted by inflation. However, there are other ways of generating such a spectrum, for example \cite{defects} by postulating the existence of a phase transition in the very early universe producing certain types of topological defects (e.g. cosmic strings or textures). The challenge for theoreticians and observers is to look for measures by means of which the predictions of these different theories can be distinguished (see e.g. Ref. \cite{andy} for some concrete suggestions).

There is no doubt that the inflationary universe scenario has led to a breakthrough in modern cosmology. It has yielded the first theory of the early universe which is able to provide a causal mechanism for the generation of density perturbations and CMB anisotropies, and which simultaneously addresses some old puzzles of Big Bang cosmology. Hopefully, the above list of provocative questions will stimulate renewed attention to some of the foundations of the scenario and will lead to research elevating the scenario to the status of a complete theory, or replacing it by a different theory.

I am grateful to the organizers of the  {\it Critical Dialogues in Cosmology Conference} (in particular Neil Turok) for setting up such a stimulating meeting. I wish to thank Andy Albrecht, Alan Guth and Bill Unruh for discussion. The author is supported in part by the U.S. Department of Energy under Grant
DE-FG0291ER40688, Task A.


\begin{thebibliography}{}

\bibitem[*]{conf} Invited contribution to the conference {\it Critical Dialogues in Cosmology}, June 24 - 27 1996, Princeton, N.J. to be publ. in the proceedings.

\bibitem{original} A. Guth, {\it Phys. Rev.} {\bf D23}, 347 (1981);\\
A. Linde, {\it Phys. Lett.} {\bf 108B}, 389 (1982);\\
A. Albrecht and P. Steinhardt, {\it Phys. Rev. Lett.} {\bf 48}, 1220 (1982);\\
A. Linde, {\it Phys. Lett.} {\bf 129B}, 177 (1983).

\bibitem{reviews} A. Linde, `Particle Physics and Inflationary Cosmology' (Harwood, Chur, 1990);\\
K. Olive, {\it Phys. Rep.} {\bf 190}, 307 (1990);\\
S. Blau and A. Guth, `Inflationary Cosmology', in `300 Years of Gravitation', ed. by S. Hawking and W. Israel (Cambridge Univ. Press, Cambridge, 1987);\\
R. Brandenberger, {\it Rev. Mod. Phys.} {\bf 57}, 1 (1985).

\bibitem{structure} V. Mukhanov and G. Chibisov, {\it JETP Lett.} {\bf 33}, 532 (1991);\\
A. Starobinsky, {\it Phys. Lett.} {\bf 117B}, 175 (1982);\\
S. Hawking, {\it Phys. Lett.} {\bf 115B}, 295 (1982);\\
A. Guth and S-Y. Pi, {\it Phys. Rev. Lett.} {\bf }, (1982);\\
J. Bardeen, P. Steinhardt and M. Turner, {\it Phys. Rev.} {\bf D28}, 679 (1983).

\bibitem{durrer} R. Durrer and J. Laukenmann, {\it Class. Quant. Grav.} {\bf 13}, 1069 (1996).

\bibitem{nonsingular} V. Mukhanov and R. Brandenberger, {\it Phys. Rev. Lett.} {\bf 68}, 1969 (1992). 

\bibitem{natural} K. Freese, J. Frieman and A. Olinto, {\it Phys. Rev. Lett.} {\bf 65}, 3233 (1990).

\bibitem{lyth} P. Binetruy and M.K. Gaillard, {\it Phys. Rev.} {\bf D34}, 3069 (1986);\\
G. Dvali, Q. Shafi and R. Schaefer, {\it Phys. Rev. Lett.} {\bf 73}, 1886 (1994);\\
D. Lyth and E. Stewart, {\it Phys. Rev. Lett.} {\bf 75}, 201 (1995).

\bibitem{ellis} G.F.R. Ellis, unpublished (1982);\\
J. Gott III, {\it Nature} {\bf 295}, 304 (1982).

\bibitem{open} J. Gott III and T. Statler, {\it Phys. Lett.} {\bf 136B}, 157 (1984);\\
M. Bucher, A. Goldhaber and N. Turok, {\it Phys. Rev.} {\bf D52}, 3314 (1995);\\
A. Linde, {\it Phys. Lett.} {\bf B351}, 99 (1995).

\bibitem{andy} J. Magueijo, A. Albrecht, D. Coulson and P. Ferreira, {\it Phys. Rev. Lett.} {\bf 76}, 2617 (1996);\\
J. Magueijo, A. Albrecht, P. Ferreira and D. Coulson, astro-ph/9605047, {\it Phys. Rev. D}, in press (1996);\\
W. Hu and M. White, astro-ph/9602020 (1996);\\
W. Hu, D. Spergel and M. White, astro-ph/9605193 (1996);\\
N. Turok, astro-ph/9604172 (1996);\\
N. Turok, astro-ph/9607109 (1996).

\bibitem{defects} T.W.B. Kibble, {\it Phys. Rep.} {\bf 67}, 183 (1980);\\
A. Vilenkin, {\it Phys. Rep.} {\bf 121}, 263 (1985);\\
A. Vilenkin and E.P.S. Shellard, `Strings and Other Topological Defects' (Cambridge Univ. Press, Cambridge, 1994);\\  
M. Hindmarsh and T.W.B. Kibble, {\it Rept. Prog. Phys.} {\bf 58}, 477 (1995);\\
N. Turok, {\it Phys. Scr.} {\bf T36}, 135 (1991);\\
R. Brandenberger, {\it Int. J. Mod. Phys.} {\bf A9}, 2117 (1994).

\end{thebibliography}
\end{document}